\documentstyle[12pt]{article}
\def\theequation{\thesection.\arabic{equation}}
\renewcommand{\baselinestretch}{1.0}
\date{ }
\def\ll{\label}

\def\c{\cite}

\def\r1{(\ref{$1})}

\def\ba{\begin{array}{c}}

\def\ea{\end{array}}

\def\l{\left}
\def\l({\left(}
\def\r){\right)}
\def\r{\right}

\def\be{\begin{equation}}
\def\bc{\begin{center}}
\def\ec{\end{center}}
\def\bit{\begin{itemize}}
\def\eit{\end{itemize}}
\def\ee{\end{equation}}
\def\ed{\end{document}}
\def\bea{\begin{eqnarray}}
\def\eea{\end{eqnarray}}
\def\efr{\end{flushright}}

\begin{document}
 
\title{{\tiny{International Journal of Modern Physics A A, Vol. 11, No. 12 (1996)
2143-2165\\
c: World Scientific Publishing Company\\}}
Quantum integrability of nonultralocal models through
Baxterisation of quantised  braided algebra
}

\author{
 Ladislav   Hlavat\'{y}
\\
{  Faculty  of  Nuclear  Sciences  and
Physical Engineering, Prague
}\footnote{
Postal   address:
B\v{r}ehov\'{a} 7, 115 19
Prague 1, Czech Republic. e-mail: hlavaty@br.fjfi.cvut.cz }
\\
and\\
 Anjan Kundu\footnote{ Permanent address: Theory Division,
Saha Institute of Nuclear Physics, AF/1 Bidhan Nagar,
700 064  Calcutta, India. e-mail: anjan.kundu@saha.ac.in}
 \\
  Physikalisches Institut der Universit\"at Bonn, 53115 Bonn,  Germany
}

\maketitle
\vskip 1 cm

\begin{abstract}
{ A scheme suitable for describing quantum nonultralocal models
including supersymmetric ones is proposed.
Braided algebras are  generalised to be used
through Baxterisation for constructing  braided
 quantum Yang--Baxter equations. Supersymmetric and some known
 nonultralocal models  are derived in the framework of the present
approach. As further applications of this scheme construction of
new quantum integrable nonultralocal models like mKdV and
anyonic supersymmetric models including deformed anyonic super algebra
are outlined.
}
\end{abstract}

\smallskip


\section {Introduction}
   \setcounter{equation}0
One of the major achievements in the theory of
  integrable systems is its extension from the classical to the quantum
domain \cite{Faddeev} and exact solution   of a number
of quantum models \cite{kul-skly}. However, inspite of this success
the
basic theory has been developed  only for a limited class of quantum models,
 known as ultralocal models  \cite{Faddeev}. For such systems
   the Lax operators at different lattice
 points $j, k ; j \neq k$  must commute, i.e.
 \be L_{2k}(v)L_{1j}(u) =L_{1j}(u)L_{2k}(v)
\ll{ulocal}\ee
and only under such constraint  the
quantum Yang-Baxter equation (QYBE)
\be R_{12}(u-v)L_{1j}(u)L_{2j}(v)=L_{2j}(v)L_{1j}(u)R_{12}(u-v), \quad
j=3, \ldots,N
\ll{rll} \ee
can be raised to the level of  monodromy matrix
\be
T_a(u) ={L}_{aN}(u){L}_{a,k-1}(u)\ldots
{L}_{a3}(u) , \quad a=1,2\ll{t}\ee
yielding the corresponding QYBE
\be R_{12}(u-v)T_{1}(u)T_{2}(v)=T_{2}(v)T_{1}(u)R_{12}(u-v)
\ll{rtt} \ee

The trace of equation (\ref{rtt})   yields a set of
commuting operators $[tr_1 T_{1}(u), tr_2 T_{2}(v)]=0$,
 ensuring the exact
  integrability of the corresponding quantum system.

On the other hand there
exists a rich class of nonultralocal  models  \cite{Maillet}
including mKdV, KdV, DNLS, chiral models
etc., which are classically
integrable, though for them
  no general
quantum theory is  available up to this date.
In the last ten years  several
 nonultralocal models were presented at the quantum
level in different contexts,
extending from integrable systems
 \cite {korepin,Nijhof} to conformal field theory (CFT) related
models \cite{babelon,alekfad,reshet,goddard}. However,
as far as we know, there was no major
effort to combine them as a  theory, except only some
indegenious proposals for
quantum maps \cite{Nijhof} and  'ultralocalisable' nonultralocal theories
\cite{Maillet91}.
 It is therefore highly desirable to formulate a quantum theory
of nonultralocal models, at least for a certain class, which would permit,
in parallel to the ultralocal case,
their systematic construction
 starting from some fundamental level.
Such a theory should include naturally the supersymmetric models, since they
are also nonultralocal in a broader sense. For achieving this nontrivial
task however, one requires clearly to generalise suitably
the basic structures like quantum algebra \cite{Drinfeld},
 Faddev-Reshetikhin-Takhtajan (FRT)
algebra \cite{FRT}, Yang-Baxter equation (YBE) etc.
With this global aim in mind we propose here some primary steps in this
 direction.

The arrangement of the paper is as follows.
In sect. 2 we  generalise the notion of quantised  braided  group
  \cite{hlavqbgrps} and its dual
algebra leading to a braided extension of the FRT algebra.
This FRT algebra is found to permit Baxterisation much in common with the
ultralocal case \cite{kundumallick}, yielding a spectral parameter dependent
braided QYBE for the Lax operator of the nonultralocal lattice models.
This is the content of sect. 3.
 In sect. 4 we  derive
 the global  QYBE for the monodromy matrix of the corresponding
 periodic models  using the associated bialgebra structure.
The quantum integrability is given by the commuting  set of operators.
Their construction from the QYBE, which  becomes nontrivial
here,   is solved in sect. 5 for
 a certain class of braidings.
In sect. 6 we find
  that the quantum supersymmetric and
nonultralocal models,  like nonabelian Toda chain
\cite{korepin}, lattice Gelfand-Dikii mapping \cite{Nijhof} as well as
CFT related models like lattice regularised Wess-Zumino-Witten-Novikov (WZWN)
 \cite{alekfad} and Toda field
theory \cite{babelon}  fit well in our proposed theory.
To illustrate further the  usefulness of this scheme  we construct in
sect. 7 new examples of quantum integrable nonultralocal models such
as quantum lattice generalisation of mKdV model, a new type of
supersymmetric model involving bosons and anyons and a quantum deformation
of the anyonic super algebra.
Sect. 8 is the concluding section.

\section { Generalised quantised braided  group and its dual algebra}
 \setcounter{equation}0
The quantised braided groups were introduced in \cite{hlavqbgrps}
combining Majid's concept of braided groups \cite{Majid2} and the FRT
formulation of quantum supergroups \cite{Liao}. The
 generators of quantised braided groups  $T=T_i^j, \  i,j \in
\{1,\ldots ,d=dimV\}$
satisfy the relation
\begin{equation}
R_{12}Z_{12} ^{-1}T_{1}Z_{12}T_{2}     =   Z_{21}   ^{-1}
T_{2}Z_{21}T_{1} R_{12}             \label{qbgreln}
\end{equation}
where the numerical matrices satisfy the system of
Yang--Baxter--type equations
\begin{equation} R_{12}R_{13}R_{23}     =
R_{23}R_{13}R_{12}
\label{rrr0}
\end{equation}
\begin{equation} Z_{12}Z_{13}Z_{23}     =
Z_{23}Z_{13}Z_{12}.
\label{zzz0}
\end{equation}
\begin{equation} R_{12}Z_{13}Z_{23}     =      Z_{23}Z_{13}R_{12},\quad
 Z_{12}Z_{13}R_{23}     =      R_{23}Z_{13}Z_{12}
\label{zzr0}
\end{equation}
The special cases are the quantum supergroups \cite{Liao}, where
$Z_{ij}^{kl}=(-)^{\hat i \hat j}\delta^k_i\delta_j^l$ and braided linear
groups \cite {Majidjmp}, where $R=Z$ and $\hat i=grad (i)$.

The dual algebra is generated by the generators
$L^{\pm} = \{L^{\pm\ j}_i\}$ that satisfy
\begin{equation}
{R}_{21}Z_{21}^{-1}L_{1}^{\epsilon}Z_{21}L_{2}^{\sigma}
= Z_{12}^{-1}L_{2}^{\sigma}Z_{12}L_{1}^{\epsilon}{R}_{21}
\ll{rpll}\end{equation}
where $(\epsilon,\sigma)=(+,+),(+,-),(-,-).$
The fundamental representation $\phi_3$ of this algebra in $V_3$ such
that
$dim V_3=dim V_2=dim V_1$ is
\be \phi_3(L_a^+)=Z_{a3}R_{3a},\ \
\phi_3(L_a^-)=Z_{a3}R_{a3}^{-1} \ \ a=1,2.\ll{frepl} \ee

The algebras (\ref{qbgreln})
 and (\ref{rpll}) can be turned into bialgebras by introducing
matrix coproducts
\be \Delta (T_i^j)=(T_i^k\otimes T_k^j), \ \
 \Delta ({L^\epsilon}_i^j)={L^\epsilon}_i^k\otimes {L^\epsilon}_k^j \ee
 provided the multiplication in the tensor product of algebras
is defined by virtue of a braid map given by matrix $Z$
as
\begin{equation}
 T_{2 }^{(k)} Z_{12}^{-1}T_{1 }^{ (j)}
= Z_{12}^{-1}T_{1 }^{(j)} Z_{12} T_{2 }^{ (k)}
Z_{12}^{-1}
\ll{tt}\end{equation}
\begin{equation}
 L_{2 }^{\epsilon_1 (k)} Z_{21}^{-1}L_{1 }^{\epsilon_2 (j)}
= Z_{21}^{-1}L_{1 }^{\epsilon_2 (j)} Z_{21} L_{2 }^{\epsilon_1 (k)}
Z_{21}^{-1}
\ll{zlzl2}\end{equation}

for $j < k$
and  ${\epsilon_1},{\epsilon_2}$ being all possible combinations
of $(+,-)$.
 Here $L^{\epsilon (j)}$,
  $T^{(j)}$ are copies of the generators $L^\epsilon, T$
 in the $j$-th factor of the
tensor product of algebras.

Now to make the theory suitable for application to nonultralocal
quantum systems we need algebras with  braiding maps  acting differently
on different pairs of factors in the tensor product.
For this purpose we
 propose a generalisation of the quantised braided  group in  the following
way.
\begin{equation}
 R_{12}  Z_{12}^{-1} T_{1}\tilde Z_{12}T_{2}     =   Z_{21}   ^{-1}
T_{2}\tilde Z_{21}T_{1}  R_{12}             \label{qbgrelng}
\end{equation}
where $Z$ can be different from $\tilde Z$.

The algebra dual to
(\ref {qbgrelng}) is
 generated by
$L^{\pm} = \{L^{\pm\ j}_i\}$ and  satisfy
\begin{equation}
{R}_{21}Z_{21}^{-1}L_{1}^{\epsilon}\tilde Z_{21}L_{2}^{\sigma}
= Z_{12}^{-1}L_{2}^{\sigma}\tilde Z_{12}L_{1}^{\epsilon}{R}_{21}
\ll{rpllg}\end{equation}
where $(\epsilon,\sigma)=(+,+),(+,-),(-,-),$ which is a braided
 generalisation of the FRT algebra \cite{FRT}.
 The possibility of introducing coproduct into the
generalised FRT algebra
(\ref{rpllg}) requires  braiding
 relations for the generators $L^{\pm}$ as
\begin{equation}
 {L_{2 }^{\epsilon_1}}^{(j+1)} Z_{21}^{-1}{L_{1 }^{\epsilon_2}}^
{(j)}
=\tilde Z_{21}^{-1}{L_{1 }^{\epsilon_2}}^{(j)}
\tilde Z_{21} {L_{2 }^{\epsilon_1}}^{(j+1)}\tilde Z_{21}^{-1}
\ll{zlzl1+}\end{equation} and

\begin{equation}
 L_{2 }^{\epsilon_1 (k)}\tilde Z_{21}^{-1}L_{1 }^{\epsilon_2 (j)}
=\tilde Z_{21}^{-1}L_{1 }^{\epsilon_2 (j)}\tilde Z_{21}
 L_{2 }^{\epsilon_1 (k)}
\tilde Z_{21}^{-1}
\ll{zlzl2+}\end{equation}
for $ k > j+1$. Note that the    relations (\ref{zlzl1+}),(\ref{zlzl2+})
are generalisation of (\ref{zlzl2}) that distinguishes between nearest and
non-nearest neighbours.

The numerical matrices $ R_{12}, \  \tilde Z_{12} , \ Z_{12}$
satisfy a system of Yang-Baxter type equations generalising (\ref{rrr0})-
(\ref{zzr0}) in the form
\begin{equation} R_{12}R_{13}R_{23}     =
R_{23}R_{13}R_{12},
\ll{rrr}
\end{equation}
\begin{equation} Z_{12}Z_{13}Z_{23}     =
Z_{23}Z_{13}Z_{12},\ll{zzz} \ee  \be  \tilde Z_{12} \tilde Z_{13} \tilde Z_{23}     =
 \tilde Z_{23} \tilde Z_{13} \tilde Z_{12}
\ll{z1z1z1}
\ee
\begin{equation} R_{12} \tilde Z_{13} \tilde Z_{23}     =
 \tilde    Z_{23} \tilde Z_{13}R_{12}
, \quad  \tilde Z_{12} \tilde Z_{13}R_{23}     =
    R_{23} \tilde Z_{13} \tilde Z_{12}
\ll{z1z1r}
\end{equation}
\begin{equation} Z_{12} \tilde Z_{13} \tilde Z_{23}     =
 \tilde Z_{23} \tilde Z_{13}Z_{12}.
, \quad \tilde Z_{12} \tilde Z_{13}Z_{23}     =
Z_{23} \tilde Z_{13} \tilde Z_{12}
\ll{z1z1z}
\ee
\begin{equation} R_{12}Z_{13}Z_{23}     =      Z_{23}Z_{13}R_{12}
, \quad Z_{12}Z_{13}R_{23}     =      R_{23}Z_{13}Z_{12}
.\ll{zzr}
\end{equation}

These equations guarantee
 the associativity of the triple product

\be
  L_{1 }^{\epsilon_1 (j_1)}(\tilde Z_{21}  L_{2 }^{\epsilon_2 (j_2)})
 (\tilde Z_{31}\tilde Z_{32}  L_{3 }^{\epsilon_3 (j_3)}) \ee
Actually the set of equations following directly from the  associativity
condition is a bit different, but they are satisfied when the system
 (\ref{rrr})-(\ref{zzr}) holds.
For example in the alternative set of equations (\ref{zzz}) does not
appear, but on the other hand one gets

\begin{equation}\tilde R_{12}\tilde R_{13}\tilde R_{23}     =
\tilde R_{23}\tilde R_{13}\tilde R_{12},
\ll{rrr1}
\end{equation}
where $\tilde R_{12}= Z_{12} R_{12}Z_{21}^{-1}$.

\section {Baxterisation of braided FRT algebra}
 \setcounter{equation}0
For framing our theory applicable to integrable systems we need to
introduce spectral parameter into the above algebraic structures.
That can be done by Baxterisation procedure \cite {kundumallick}.

Introducing $R^+_{12}=R_{21}$ and
 $R^-=P(R^+)^{-1}P$  with $P$ being the permutation matrix, we rewrite the
braided FRT algebra (\ref {rpllg}) in the following form, convenient for the
Baxterisation.
\begin{equation}
{R}^+_{12}Z_{21}^{-1}L_{1}^{\epsilon}\tilde Z_{21}L_{2}^{\sigma}
= Z_{12}^{-1}L_{2}^{\sigma}\tilde Z_{12}L_{1}^{\epsilon}{R}^+_{12}
\ll{rpll+}\end{equation}
and
\begin{equation}
{R}^-_{12}Z_{21}^{-1}L_{1}^{\sigma}\tilde Z_{21}L_{2}^{\epsilon}
= Z_{12}^{-1}L_{2}^{\epsilon}\tilde Z_{12}L_{1}^{\sigma}{R}^-_{12}
\ll{rpll-}\end{equation}
where $(\epsilon,\sigma)=(+,+),(+,-),(-,-).$

 Now to build spectral paremeter dependent  Lax operator and the quantum
$R$-matrix satsifying the corresponding QYBE we set \cite{kundumallick}
\be R_{12}(u)=e^uR^+_{12}-e^{-u}R^-_{12} \ll{baxr} \ee
and
 \be L_a(u)=e^uL_a^+-e^{-u}L_a^- , \quad a=1,2    \ll{baxl}
\ee
and observe that the Baxterisation goes through, only when
 apart from (\ref{rpll+}) and  (\ref {rpll-}) the following extra relation

\begin{equation}
{R}^-_{12}Z_{21}^{-1}L_{1}^{+}\tilde Z_{21}L_{2}^{-}
-{R}^+_{12}Z_{21}^{-1}L_{1}^{-}\tilde Z_{21}L_{2}^{+}
= Z_{12}^{-1}L_{2}^{-}\tilde Z_{12}L_{1}^{+}{R}^-_{12}
- Z_{12}^{-1}L_{2}^{+}\tilde Z_{12}L_{1}^{-}{R}^+_{12}
\ll{Baxcond}\end{equation}
holds. However  in analogy with \cite{ kundumallick}
 one finds that the above relation
is satisfied under the
 Hecke condition
\be R^+-R^-=cP \ll{Hecke} \ee
on $R^{\pm}$ matrices, which leads finally to the spectral parameter
dependent braided equation
\begin{equation}
{R}_{12}(u-v)Z_{21}^{-1}L_{1}(u)\tilde Z_{21}L_{2}(v)
= Z_{12}^{-1}L_{2}(v) \tilde Z_{12}L_{1}(u){R}_{12}(u-v)
\ll{rzlzl}\end{equation}
 with $R_{12}(u)$ satisfying the standard YBE
\be R_{12}(u-v)R_{13}(u)R_{23}(v)=
 R_{23}(v)R_{13}(u)R_{12}(u-v)\ll{ybe} \ee
along with the equations

\begin{equation} R_{21}(u)Z_{13}Z_{23}     =
Z_{23}Z_{13}R_{21}(u)
\label{ruzz}
\end{equation}
\begin{equation} Z_{12}Z_{13}R_{32}(u)=   R_{32}(u)Z_{13}Z_{12}
\label{zzru}
\end{equation}
and similar equations for $\tilde Z_{12}$.
We would like to mention here that the Baxterised solutions  (\ref{baxr})
of YBE satisfy automatically the unitarity condition
\[R(u)PR(-u)P=(c^2-4\sinh^2 u){\bf 1}. \]

\section {Braided QYBE's}
 \setcounter{equation}0
For   application of the above formulated generalised braided algebras
to
integrable sytems we define the Lax operators
$L_j(u)$ of the associated discrete model
which  satisfy  (\ref{rzlzl})
 as the braided local QYBE
\begin{equation}
{R}_{12}(u-v)Z_{21}^{-1}L_{1j}(u)\tilde Z_{21}L_{2j}(v)
= Z_{12}^{-1}L_{2j}(v) \tilde Z_{12}L_{1j}(u){R}_{12}(u-v).
\ll{bqybel}\end{equation}
For the construction of models
we need a  monodromy matrix $T^{[k,j]}_{a}(u), \ k>j$
defined as a product  of the Lax operators
\be T_a^{[k,j]}(u) ={L}_{ak}(u){L}_{a,k-1}(u)\ldots
{L}_{aj}(u)\ll{Tkju}\ee
acting in the space
${\cal H} \equiv V_k\otimes V_{k-1}\otimes\ldots\otimes V_j$
and satisfying the same relation as the $L_{aj}(u)$ operators.
This can be achieved   if the Lax operators
along with (\ref{bqybel}), satisfy also the
Baxterised forms of braiding relations (\ref{zlzl1+}),(\ref{zlzl2+}):
\begin{equation}
 L_{2 j+1}(v)Z_{21}^{-1}L_{1 j}(u)
=\tilde Z_{21}^{-1}L_{1 j}(u)\tilde Z_{21} L_{2 j+1}(v)\tilde Z_{21}^{-1}
\ll{zlzl1u}\end{equation} and

\begin{equation}
 L_{2 k}(v)\tilde Z_{21}^{-1}L_{1 j}(u)
=\tilde Z_{21}^{-1}L_{1 j}(u)\tilde Z_{21} L_{2 k}(v)\tilde Z_{21}^{-1}
\ll{zlzl2u}\end{equation}
for $k>j+1$.

Using now the relations (\ref{bqybel}),
 (\ref{zlzl1u}) and  (\ref{zlzl2u}) it is straightforward to show that
  the monodromy matrix satisfies the equation

\begin{equation}
{R}_{12}(u-v)Z_{21}^{-1} T_{1}^{[k,j]}(u)\tilde Z_{21} T^{[k,j]}_{2}(v)
= Z_{12}^{-1} T^{[k,j]}_{2}(v) \tilde Z_{12} T^{[k,j]}_{1}(u){R}_{12}(u-v)
\ll{rztzt}\end{equation}
Note that the ultralocal condition (\ref{ulocal})
 is generalised here to the braiding
relations (\ref{zlzl1u}) and (\ref{zlzl2u}), which carry the main feature of
nonultralocality. It can be seen from (\ref{zlzl1u}),(\ref{zlzl2u}) that
  the present theory is designed  to cover  models
with nonultralocal commutation relations  different for nearest and
nonnearest neighbours.

We intend to incorporate now the  periodic boundary condition
$ L_{aj+P}(u)= L_{aj}(u)$  to approach closer to
the physical models and consider the global monodromy matrix $ T_a(u) \equiv
  T^{[N,3]}_{a}(u)$ for the closed chain $ [N,3] $.
 Observe however that for deriving
equation for the monodromy matrix $T_a$  one requires
 along with  (\ref{zlzl1u}) and
 (\ref{zlzl2u}) also the braiding relation like
\begin{equation}
 L_{2 j-1}(v)\tilde Z_{21}^{-1}L_{1 j}(u)
=\tilde Z_{21}^{-1}L_{1 j}(u) Z_{12}^{-1} L_{2 j-1}(v)\tilde Z_{21}^{-1}
\ll{zlzl3}\end{equation} the compatibility of which with (\ref {zlzl1u})
demands the constraint
\be \tilde Z_{12} \tilde Z_{21}=\bf {1}
.\ll{zz}\end{equation}
Since $ L_{3-1}(u)= L_{N}(u),$ using (\ref{zlzl3}) for $j=3$
together with (\ref{zlzl1u}), (\ref{zlzl2u}) we obtain
finally the braided QYBE for the monodromy matrix as
\begin{equation}
{R}_{12}(u-v)Z_{21}^{-1} T_{1}(u)Z_{12}^{-1} T_{2}(v)
= Z_{12}^{-1} T_{2}(v) Z_{21}^{-1} T_{1}(u){R}_{12}(u-v)
\ll{bqybet}\end{equation}

\section{Trace identity and quantum integrability}
 \setcounter{equation}0
For ensuring quantum integrability we need a
commuting set of conserved quantities.
By taking the trace of (\ref{bqybet}) we arrive at
\begin{equation}
tr_{12}\left(Z_{21}^{-1} T_{1}(u)Z_{12}^{-1} T_{2}(v)\right)
=tr_{12}\left( Z_{12}^{-1} T_{2}(v) Z_{21}^{-1} T_{1}(u)\right)
\ll{trace}\end{equation}
 For $Z=1$, i.e. for  unbraided or equivalently ultralocal
 case, the above trace identity is trivially factorised yielding the
required commuting set of operators.
However for general $Z$ the trace factorisation problem, namely reducing
 the trace identity (\ref{trace})
to the form
\[tr_{1}(\tilde T_{1}(u) )tr_{2}(\tilde T_{2}(v) )=
tr_{2}(\tilde T_{2}(v) )tr_{1}( \tilde T_{1}(u) ) \]
becomes nontrivial.
For solving  this problem we propose
the following  factorisation procedures.
\subsection {  Factorisation by k-trace}
This procedure is
 based on
the generalisation of
 Sklyanin's approach to  reflection--type algebras \cite{Sklyanin}.
It was shown in \cite{hlavopsc} that if $T_i^j(u),K_i^j(u)$
are generators of the algebra satisfying quadratic relations
\be T_1(u)K_2(v) = K_2(v)T_1(u)\ll{mkkm} \ee
\be A_{12}(u,v)T_1(u)B_{12}(u,v)T_2(v)=
T_2(v)C_{12}(u,v)T_1(u)D_{12}(u,v),
\ll{ambm} \ee
\be A_{12}^{-T}(u,v)K_1^{t_1}(u)\tilde B_{12}(u,v)
K_2^{t_2}(v)=K_2^{t_2}(v)
\tilde C_{12}(u,v)K_1^{t_1}(u)D_{12}^{-T}(u,v)
\ll{akbk} \ee
where
\be \tilde B_{12}(u,v)={({(B_{12}^{t_1}(u,v))}^{-1})}^{t_2},\quad
 \tilde C_{12}(u,v)={({(C_{12}^{t_2}(u,v))}^{-1})}^{t_1}
\ll{btct} \ee
with  $A, B, C, D \  $ numerical matrix functions,
then the 'k-trace' $ \ t(u)={K}_i^j(u)T_j^i(u)=tr[K(u)T(u)]$
 forms a commutative
subalgebra
\[ [t(u),t(v)] = 0. \]

In our case of (\ref{bqybet})
\[A(u,v)= ZR(u-v)PZ^{-1}P,\ \ B=Z^{-1},\ \ C=PZ^{-}P,\ \ D(u,v)=R(u-v).\]
Therefore we have to find a set of operators $K_i^j(u)$ on $\cal
H$ commuting with $T^j_i(u)$ and satisfying (\ref{akbk})
.
The simplest possibility is $K_i^j(u)=k_i^j(u){\bf 1}_{\cal H}$
where $k^j_i(u)$ are elements of the numerical matrix $k(u)$
satisfying
\be k_1(u)\hat Z_{21} k_2(v)Z_{12}R_{12}(u-v)=
R_{12}(u-v)k_2(v)\hat Z_{12}k_1(u)Z_{21}. \ll{rzkzk} \ee
where
\be \hat Z={({((Z^{-1})^{t_1})}^{-1})}^{t_1}
.\ee
It is true that equation (\ref{rzkzk})  determining the
condition for factorisability of  traces
is rather difficult to solve in the
general case. Nevertheless,
in our attempt to   simplify  this equation
we observe interestingly that, for some special ans\"atze of  $Z$
one may obtain  $\hat Z= Z$. These ans\"atze for $Z$ are
\\  a){  {\it the diagonal form \  \ }}
\\   b){ {\it \ factorised form}: \ \ $ Z=
A\otimes B$}
 \\c)   {\it specifically chosen form}:
 \be  \quad \ \  Z_{12}={\bf 1}+\sum_iA_i\otimes B_i, \
\ , \ \ (B_i)^2=0, \ll{case2} \ee
\noindent d) {\it $Z$ coinciding with $R_q$}:
 \be  \quad \ \  Z_{12}=  R^+_{q21}
, \ll{ex4} \ee
\noindent  e) {\it $Z$ expressed as}
 \be  \quad \ \  Z_{12}=
e^{h\sum_i A_i
\otimes B_i}. \ll{ex5} \ee
with arbitrary invertible matrices $A_i, B_i$.
The related reduction  $\hat Z= Z$ allows to find now
 some particular  solutions of the factorisabilty equation (\ref{rzkzk}) as
$ k={\bf 1}$ under the condition \be {R}_{12}(u)Z_{12}Z_{21}
=Z_{21}Z_{12}{R}_{12}(u) \ \ll{sym2}\ee
causing the
ordinary trace of $T(u)$ to produce the set of commuting operators.

Let us now have a closer look   into  particular cases
of the above  proposal, while  another possibility of  factorisation
 will be given below. Finally  we will  show that the known nonultralocal
as well as supersymmetric  models fit well into these solutions.
\\ \\
1.
 {\it  Let $Z$ be a diagonal matrix ( case a) above)}:
  \be Z_{12}=\sum_{\alpha,\beta}
g_{\alpha \beta}
e_{\alpha \alpha}\otimes e_{\beta \beta}
, \quad \mbox{where} \ \
(e_{\alpha \beta})_{\gamma\delta}
 =\delta_{\alpha\gamma} \delta_{ \beta\delta }.
\ll{case4}\ee
 If
\be
g_{\alpha \beta}g_{ \beta
\alpha}=1 ,\ll{gg=1}\ee
then  (\ref{sym2}) is trivially satisfied giving $k=1$, i.e. commutativity
of  $ \ tr T$.
We  show below that the condition (\ref{gg=1})
 holds in the
supersymmetric case and
one can construct SUSY invariant integrable models
with the  solution (\ref{case4})
and  $\tilde Z=Z$.   On the other hand
when
 $g_{\alpha \beta}=g_{\beta \alpha }$ is symmetric, it commutes with
 the $R$-matrix of the class (\ref{R(u)}) and hence
 leads to  the  'k-trace'  with  arbitrary $k$ commuting with such
 $R$ and $Z$-matrices.
Such a possibility will be applied in sect. 7.2-3 for
 constructing anyonic quantum integrable models.
\\ \\
\quad 2. {\it  Let $ Z_{12}$ be in the factorised form
 $ Z_{12}^{-1}= A_1B_2$
 (case b) above)}.\\The equation (\ref{rzkzk}) in this case
can be solved easily for arbitrary
$R$
yielding $ \ k=AB $. Therefore as follows from the above result,
 the 'k-trace' or $ \ tr \tilde T= tr( A B T)$
would  generate the
commuting conserved quantities. This conclusion can also be
  confirmed by
inserting  the factorised form of $Z$  directly in
our trace formula (\ref{trace}).
\\ \\
\quad 3. {\it Let $ Z_{12}$ exhibits the  symmetry
   (\ref{sym2})  and
  restricted as  in the case c) above}.\\
We see that in this case $k=1$ and using (\ref{sym2}) the braided QYBE's
(\ref{bqybel}), (\ref{bqybet})   may
 be  transformed to
 \begin{equation}
{R}_{12}(u-v)Z_{12}L_{1j}(u)\tilde Z_{21}L_{2j}(v)
= Z_{21}L_{2j}(v) \tilde Z_{12}L_{1j}(u){R}_{12}(u-v)
\ll{bqybel2}\end{equation}
and
\begin{equation}
{R}_{12}(u-v)Z_{12}T_{1}(u) Z_{12}^{-1}T_{2}(v)
= Z_{21}T_{2}(v)  Z_{21}^{-1}T_{1}(u){R}_{12}(u-v)
.\ll{bqybet2}\end{equation}

Due to  $k=1$,
  $tr T$ will now exhibit commuting relations, which  we have also checked
independently
by taking trace of (\ref{bqybet2}) and using the form   (\ref{case2}).
 A concrete realisation of this case  will be shown below through
an example.
 \\ \\ \quad
 4. {\it Let  $Z_{12}=  R^+_{q21}$  (case d) above)} \\
Using the Baxterisation formula (\ref{baxr}) and the relation
$  R^-_{12}=(R^+_{21})^{-1}$ we may check that this
 is a  solution  of   (\ref{sym2}) for $Z$ so that  $k=1$, which
ensures the trace factorisability. Moreover
  the same  choice for $Z$
satisfies  also   the eqns.
(\ref{zzz}) and
(\ref{zzr}) for $ R_{12} = R^+_{q21}   $
and consequently, it can serve as a right
candidate for constructing nonultralocal quantum integrable models.

To see how   the relations are converted in this special case, which will
be considered in the next section,
we use the relation $ R^+_{q12}= (R^+_{t12})^{-1}=
R^-_{t21}$  and $ Z^{-1}_{21}= R^+_{t12}$ with $t=q^{-1}$
and  rewrite  (\ref{rpll+}) as
\[
({R}^+_{t12})^{-1}{R}^+_{t12}L^\epsilon_{1j} \tilde Z_{21} L_{2j}^\sigma
= ({R}^-_{t12})^{-1}L_{2j}^\sigma \tilde Z_{12}L_{1j}^\epsilon
({R}^+_{t12})^{-1}\] or
\be {R}^-_{t12}L_{1j}^\epsilon \tilde Z_{21} L_{2j}^\sigma
{R}^+_{t12}=L_{2j}^\sigma \tilde Z_{12}L_{1j}^\epsilon
\ll{fadcurrent0}\ee
 and
the braiding relations
(\ref{zlzl1+}-\ref{zlzl2+})
as
\begin{equation}
 L_{2 j+1}^{\epsilon_1}
{R}^+_{t12}L_{1 j}^{\epsilon_2}
=\tilde Z_{21}^{-1}L_{1 j}^{\epsilon_2}\tilde Z_{21}L_{2 j+1}^{\epsilon_1}
\tilde Z_{21}^{-1}
,\quad
L_{2 k}^{\epsilon_1} \tilde Z_{21}^{-1}L_{1 j}^{\epsilon_2}
=\tilde Z_{21}^{-1}L_{1 j}^{\epsilon_2}
\tilde Z_{21} L_{2 k}^{\epsilon_1}\tilde Z_{21}^{-1}
\ll{fadcurrent2}\end{equation}
for $k>j+1$.

The expression for  $ \tilde Z\neq 1 $ must be obtained by solving
its relevant equations.
 For example, the simplest choice in this case may be
   $ \tilde Z= 1 $ or  $ \tilde Z_{12}= Z_{12}=  R^+_{21} $.
\\ \\ \quad
5.
 {\it Let  $Z_{12}   = e^{-\frac {i} {2}h \sum_i H_i
\otimes H_i},$ where $H_i$'s are commuting matrices
  (case e) above with $A=B$)}\\
In this case $ Z_{12}= Z_{21}=\hat  Z_{12}$ and
for $ k_{i}$ commuting with $ Z_{12}$
 the equation
(\ref{rzkzk}) reduces to
\be
[ R(u-v), C(u,v)]=
0, \ \ C(u,v)= Z^2_{12}k_1(u)k_2(v). \ll{rzkzk11} \ee

If now we limit ourselves to the  triginometrix $R(u-v)$-matrix (\ref{R(u)})
and consider $H_i=e_{ii}-e_{i+1 i+1}$, then we may
 extract a solution  $k = e^{-\frac {i} {2}h \sum_i H^2_i
} $, since $ C_{12}
=Z^2_{12}k_1k_2= e^{-\frac {i} {2}h \sum_i (H_{i1}+ H_{i2})^2}$
and $ H_{i1}+ H_{i2}$ commutes with such $R(u-v)$-matrix.

It is worth noting that defining $\tilde L_{aj}=k_a^{-\frac {1} {2}}L_{aj}$
we can eliminate in this case the  $Z$ dependence from the braided QYBE
(\ref{bqybel}) yielding
\be
{R}_{12}(u-v)\tilde L_{1j}(u) \tilde Z_{21}\tilde L_{2j}(v)
=\tilde L_{2j}(v)\tilde Z_{12}\tilde L_{1j}(u)
{R}_{12}(u-v)
\ll{babcurrent10}\end{equation}
and similar simplifications can be achieved in (\ref{rztzt}),(\ref{bqybet}).

We will see below how some known nonultralocal models
are realised as  cases 4 and 5.
There are however other models that
  go beyond  the  above ans\"atze, for which
we present here another factorisation procedure.
\subsection {Factorisation by 'gauge transformation'}

{  Let
  $\hat l$ is a matrix of operators
 acting in the Hilbert space  satisfying
the relation}
\be [{R}_{12}(u),\hat l_{1 }\hat l_{2}]=0 \ \ \mbox {}\ \
[\hat l_{1 },\hat l_{2}]=0
\ll{sym3}\ee
and
 \be Z_{21}^{-1}T_1=\hat l_2T_1\hat l_2^{-1}
.\ll{case3}\ee
We show  that it gives another possibility of   trace
factorisation. Starting from
(\ref{bqybet}) and using (\ref{case3}) and (\ref{sym3})
we   get
\begin{equation}
R_{12}(u-v)\hat l_{1}^{-1} T_{1}(u)\hat l_1
\hat l_{2}^{-1} T_{2}(v)\hat l_2
=\hat l_{2}^{-1} T_{2}(v)\hat l_2
\hat l_{1}^{-1} T_{1}(u)\hat l_1
R_{12}(u-v)
.\ll{trace3}\end{equation}
Taking trace of (\ref{trace3}) results in
 the required commutation relation $[tr \tilde T(u), tr \tilde T(v)]
=0$, where
$\tilde T(u)=\hat l^{-1} T(u)\hat l$

\section {Quantum nonultralocal models }
 \setcounter{equation}0
For application to integrable models we may take any of the
above ans\"atze for $Z$ ensuring  the trace factorisability, which is
a basic requirement of integrability  and look for different solutions
of   (\ref{zzz}) and (\ref{ruzz})--(\ref{zzru})  for a given
 $R$-matrix. It is important to note that the
factorisation procedure
does not touch the $\tilde Z$-matrix allowing it to be any
arbitrary solutions
of (\ref{z1z1z1})--(\ref{z1z1z}) with additional constraint (\ref{zz})
in case of periodic chains.
We show below  that the nonultralocal quantum
models
 including   suppersymmetric
 models
can be explained   using the
present framework
 for the simplest choice  $\tilde Z=1$ or $
\tilde Z= Z$. These  examples are
 associated with different  solutions
for $Z$ and $R$-matices. It is curious to note that all
 examples known to us, including those related to CFT,
are in agreement with above forms of $Z$ leading to  trace
factorisability.

 Since Hecke condition underlies
our construction we restrict to the $R^{\pm}_q$ matrices  of \\
  $ U_q(sl(n | m))\quad $
with $q=e^{i\eta}$ leading to the trigonometric solution
 \be R_{12}(u)=\sum_{\alpha,\beta}\left(\sin u
e_{\alpha \alpha}\otimes e_{\beta \beta} +\epsilon_\alpha
\sin( u+\epsilon_\alpha\eta)
e_{\alpha \alpha}\otimes e_{\alpha \alpha}
+\sin \eta
e_{\alpha \beta}\otimes e_{\beta \alpha}\right)\ll{R(u)}\ee
where  $\epsilon_\alpha=\pm 1$  are obtained
from the Baxterisation of 'standard' or 'nonstandard' solutions, respectively.
At $q \rightarrow 1$  (\ref{R(u)}) reduces to the  rational solution of
YBE.
Note  that this type of $R$-matrix was proposed  by Perk and Schultz \cite
{Perkschultz} in connection with integrable statistical models.
Such $R$-matrices exhibit a symmetry
  \be [R_{12}(u),s_1s_2]
=0, \quad s=
\sum_{\alpha}
e_{\alpha \alpha}\phi_\alpha
 \ll{strace}\ee
Therefore if $s$ commutes with $Z$, along with the usual trace
$tr T$ one  also
gets the commutativity of 's-trace'  $tr(sT)$ . This
property  is used for example,
in the construction of SUSY models for physical reasons.

We look now into some concrete  examples of quantum nonultralocal models
known in the present day literature, which were proposed over the years in
various contexts and show that
in the framework of the present theory based on the quantised braided
 algebra it is possible to describe them in an unifying way.

\subsection
 { Supersymmetric models }
  Quantum integrable supersymmetric
theory \cite{SUSY,SUSYq}
 is possible to formulate in a convenient alternative way
 expressing the gradings in a matrix form as proposed in  \cite{Liao}. We
  observe
that such a formulation can be   reproduced nicely within our general framework
 and
corresponds to  {\it case} 1 in the previous section,
where now $Z=\tilde Z=\eta$ and
$\eta$ is in the form (\ref{case4})
 with $g_{\alpha\beta}=(-1)^{\hat \alpha \hat
 \beta}$. Here $\hat \alpha$   describes the supersymmetric grading with
 $\hat \alpha =0(1)$ depending on the even (odd)-ness of the
index. The quantum integrability follows from the commuting traces
 due to (\ref{sym2}), though however for physical reasons 'supertraces' are
usually used for such models, which may be obtained from the 's-trace'
({\ref{strace}) by choosing $\phi_\alpha=(-1)^{\hat \alpha}.$
The corresponding super QYBE's for these
SUSY models should be derived from
 (\ref{bqybel}),(\ref{bqybet}) as

\begin{equation}
{R}_{12}(u-v)\eta_{12}L_{1j}(u) \eta_{12}L_{2j}(v)
= \eta_{12}L_{2j}(v) \eta_{12}L_{1j}(u){R}_{12}(u-v)
\ll{bqybels}\end{equation}
\begin{equation}
{R}_{12}(u-v)\eta_{12} T_{1}(u)\eta_{12} T_{2}(v)
= \eta_{12} T_{2}(v) \eta_{12} T_{1}(u){R}_{12}(u-v)
\ll{bqybets}\end{equation}
while from the nonultralocal braiding relations
 (\ref{zlzl1u}) and (\ref{zlzl2u})
one obtains the supercommutation relations as

\begin{equation}
 L_{2 k}(v)\eta_{12}L_{1 j}(u)
=\eta_{12}L_{1 j}(u)\eta_{12} L_{2 k}(v)\eta_{12}
\ll{zlzl24}\end{equation}
for $k\neq j$.
In order that these equations may be compared with the similar ones appearing
in the literature on SUSY models
 \cite{SUSY,SUSYq} we express (\ref{bqybels})
 in the matrix element form

\begin{equation}
{R}_{{a_1}a_2}^{{b_1}b_2}(u-v)(L_{b_1}^{c_1}(u))_j  (L_{b_2}^{c_2}(v))_j
(-1)^{\hat b_2 (\hat b_1+\hat c_1)}
=(-1)^{\hat a_1 (\hat a_2+\hat b_2)}(L_{a_2}^{b_2}(v))_j  (L_{a_1}^{b_1}(u))_j
{R}_{{b_1}b_2}^{{c_1}c_2}(u-v)
\ll{bqybelse}\end{equation}
and the braiding relation   (\ref{zlzl24}) as
\be
(L_{a_2}^{b_2}(u))_k  (L_{a_1}^{b_1}(v))_j
=(-1)^{(\hat a_1 +\hat b_1)(\hat a_2 +\hat b_2)}
(L_{a_1}^{b_1}(u))_j  (L_{a_2}^{b_2}(v))_k
.\ee

 Note that in contrast to nonultralocal models discussed below
 different braiding relations coincide here showing 'long range
nonultralocality' of SUSY models. Choosing the $R$-matrix as rational
or trigonometric solutions we can derive the results of \cite{SUSY} and
\cite{SUSYq}, respectively.

An immediate generalisation of SUSY models to their braided analogs
can be done by putting $\tilde Z=Z$, where $Z$ is a solution of
(\ref{zzz}), (\ref{ruzz}) and
(\ref{zzru}) that enables trace factorisation like
for example  in the cases  discussed in the previous section.

We stress again  that for
 supersymmetric models or their braiding analogs we
 limit ourselves
 to     $\tilde Z=Z$  that exhibits 'homogeneous' long range
nonultralocality, whereas for inclusion of the
 following models with only nearest-neihgbour
 nonultralocality  we need $Z\neq \tilde Z$.

\subsection
 { Nonabelian Toda chain }

This is a one-dimensional evolution model on a periodic lattice
of $N$ sites described by
the nonabelian matrix-valued operator
$g_k \in GL(n)$. The model, representing  a discrete
analog of principal chiral field, was  set into the Yang-Baxter
formalism in \cite{korepin}. The Lax operator
of this  nonultralocal
 quantum model
was presented as
\be
L_{k}(\lambda) = \left( \begin{array}{c}
\lambda - A_k
\qquad  \ \ -B_{k-1}
 \\ I
\qquad \ 0
         \end{array}   \right), \ \ A_k=\dot{g}_k g_k^{-1}, \ \  B_k= g_{k+1}
 g_k^{-1}
\ll{Lkor}\ee
and the associated
 $R$-matrix is
 the rational solution of YBE $R(\lambda)=ih P-\lambda$
 with $P= \Pi\otimes \pi  ,$
where $\Pi$
is a $4 \times 4$ and $\pi$   a $\ n^2 \times n^2$ permutation matrix.
We observe that this model can  be described nicely by our formalism
by choosing  the related  $Z,\tilde Z $
 matrices as  $\tilde Z=1$  and $Z_{12}=
 {\bf 1} + i { h }(
 e_{22}\otimes e_{12})\otimes \pi$. Moreover we notice that this is an example
of the  case discussed in subsection 6.2, since by choosing
 $\hat l= diag({\bf I},g_N)$ and
 using the explicit form (\ref{Lkor}) of $L_i(u),$
 one can show  \cite{korepin}
 that $Z_{21}^{-1}L_{1N}=\hat l_2L_{1N}\hat l_2^{-1}
$ leading to the requirement (\ref{case3}):
$ Z_{21}^{-1}T_1=\hat l_2T_1\hat l_2^{-1}$. The  validity of other conditions
(\ref{sym3}) can also be readily checked.
Therefore the quantum integrability of the model is evident from
trace identity  (\ref{trace3}).
 With this input one finds easily that
our main formulas
(\ref{bqybel}) and
(\ref{bqybet}) recover exactly the QYBE's of \cite{korepin}, while
 the nonultralocal braiding relations should be given by
\begin{equation}
 L_{2 j+1}Z_{21}^{-1}L_{1 j}(u)
=L_{1 j}(v)L_{2 j+1}(v)
,\quad
L_{2 k}(v)L_{1 j}(u)
=L_{1 j}(u) L_{2 k}(v)
\ll{korepin2}\end{equation}
for $k>j+1$.

Now we look into  lattice regularised models related to  the conformal
field theory, which were
proposed to describe current and exchange
algebras in the WZWN  \cite{alekfad} and in Toda field theory \cite{babelon}.
 Since the approach of \cite{alekfad} and \cite{babelon}
is  not restricted by the demand of integrability, we may
limit ourselves to the spectral parameter independent equations
and in principle consider  $Z$ without having
the  trace factorisation property discussed in sect. 5.

\subsection { Quantum symmetry in WZWN model}

The WZWN model is described by the unitary unimodular matrix-valued field
 $g(x,t): M \rightarrow su(2), \ \ M=S^1 \times R^1$.
Defining   the chiral left current as
$L= \frac {1} {2} (J_0+J_1)$, where $J_\mu= \partial_\mu gg^{-1},$
one gets the well known current algebra relation in the form of Poisson
bracket
\[ \{L_1(x),L_2(y)\}= \frac {\gamma} {2} [C,L_1(x)-L_2(y)]\delta(x-y)
+ \gamma C \delta'(x-y) \]
where $C_{12}= 2P_{12}-1$,
  giving the Kac-Moody algebra with the central charge defined by the
coupling constant $\gamma$.
Introducing $\partial_xu = L u$ one can write the monodromy as
$M_L=u(x)^{-1}u(x+2\pi)$ and the similar relations for the right current.
With the aim of unravelling the quantum group structure in WZWN model
Faddeev and his collaborators \cite{alekfad} formulated a discrete and quantum
version of the current algebra, where $L_j$ now
stands  for the lattice regularised
current in the WZWN model and  the chiral component
becomes $u(x)\rightarrow u_j$.
The nonultralocal current algebra  at this discrete level was shown
\cite{alekfad} to be equivalent to the commutation relations
\be
{R}^+_{t12}L_{1j} L_{2j}
{R}^-_{t12}=L_{2j}L_{1j}
\ll{fadcurrent1}\end{equation}

\begin{equation}
 L_{2 j+1}
{R}^+_{t12}L_{1 j}
=L_{1 j}L_{2 j+1}
,\quad
L_{2 k}L_{1 j}
=L_{1 j} L_{2 k}
\ll{fadcurrent21}\end{equation}
for $k>j+1$, while the algebraic relations for the chiral components
were represented by

\be
{R}^+_{t12}u_{1j} u_{2j}
{R}^-_{t12}=u_{2j}u_{1j}
\ll{fadchiral}\end{equation}
\be
u_{1j} u_{2n}
{R}^+_{t12}=u_{2n}u_{1j}
\ll{fadchiral2}\end{equation}
for $j>n$
and for the monodromy as
\be
M_{L1}
({R}^-_{t12})^{-1} M_{L2}
{R}^-_{t12}=
({R}^+_{t12})^{-1}M_{L2}
{R}^+_{t12}M_{L1}
\ll{fadmonodromy}\end{equation}

We show that all these results can  be reproduced from
the general scheme presented here. For this  we
set $\tilde Z=1$ and $Z_{12}=R_{t12}^-$ and
note that this is the solution considered above as {\it case} 4
and agrees with our ansatz d).
Further we identify
$L_{aj}$ of \cite{alekfad} with our $L_{aj}^+$
and define the analogous
'monodromy' matrices $ T^{[j,k]}$ and $T$ as in (\ref{Tkju}) and (\ref{t}).
From (\ref{fadcurrent0}) with   $\tilde Z=1$ we then get

\be {R}^-_{t12}L_{1j} L_{2j}
{R}^+_{t12}=L_{2j}L_{1j}
\ll{fadcurrent01}\ee
the equivalence of which with the current algebra relation
(\ref{fadcurrent1}) is evident.
 Similarly
our braiding relations
(\ref{fadcurrent2})
recover the nonultralocal relation  at different points  (\ref{fadcurrent21}).
 Identifying
 $u_j=T^{[j,3]}$ and
 $M$  with $T$
we obtain the commutation relations for
 chiral components (\ref{fadchiral})
as well as for the monodromy  (\ref{fadmonodromy}).

Let us remark that in this specific case we can derive also the commutation
relation between $T^{[j,k]}$ and $T^{[n,k]}$:

\begin{equation}
\tilde Z_{12} T_{1}^{[j,k]}\tilde Z_{21} T^{[n,k]}_{2}{R}_{t12}^+
=  T^{[n,k]}_{2} \tilde Z_{12} T^{[j,k]}_{1}
\ll{rztzt+}\end{equation}
for $j>n$. Putting $\tilde Z=1$ in the above equation
we get the exchange relations for the chiral components (\ref{fadchiral2}).

\subsection
 { Quantum group structure in Coulomb gas picture of  CFT}

The Coulomb gas picture of CFT is based on the Drinfeld-Sokolov linear
system   \cite{babelon}
$\ \ \partial Q= L(x)Q,\ $ with the Lax operator \[  L(x)= P(x)- {\cal E}_+,
 \quad  {\cal E}_+=\sum_{\alpha\ simple\ roots} E_\alpha .\] The periodic field
$P(x)$ satisfies the Poisson bracket relation
\[ \{ P(x) \otimes, P(y) \}= \delta'(x-y) \sum_i H_i \otimes H_i \]
causing the nonultralocal nature of the model. In
  \cite{babelon} a lattice regularised description of the
 algebraic structures was given
at the quantum level, where the commutation relation of the discretised
Lax operator was presented in the form
\be
{R}^+_{12}\tilde L_{1j}\tilde L_{2j}
=\tilde L_{2j}\tilde L_{1j}
{R}^+_{12}
\ll{babcurrent1}\end{equation}
and

\begin{equation}
\tilde L_{2 j+1}
{A}_{12}\tilde L_{1 j}
=\tilde L_{1 j} \tilde L_{2 j+1}
,\quad
\tilde L_{2 k}\tilde L_{1 j}
=\tilde L_{1 j}\tilde L_{2 k}, \ \ k>j+1
\ll{babcurrent2}\end{equation}
where
${A}_{12}=q^{\sum_i H_i
\otimes H_i}, \ \
q= e^{\frac {i} {2}h
}$. From these basic
starting relations
 the algebras of monodromies were found
as
\be
{R}^+_{12} Q_{1j}Q_{2j}
=Q_{2j}Q_{1j}
{R}^+_{12}
\ll{babmono1}\end{equation}
for  $Q_j=\tilde L_jB \tilde L_{j-1}B \ldots B\tilde L_3$
 with $
B_a=q^{\frac {1} {2} \sum_\alpha H^2_\alpha}$ and
for the periodic lattice of $N$ sites
 with monodromy $S=\tilde L_N
B \tilde L_{N-1}B \ldots B\tilde L_3$
as
\be
{R}^+_{12} S_{1}A_{12}S_{2}
=S_{2}A_{12}S_{1}
{R}^+_{12}
\ll{babmono2}\end{equation}
For relating this structure with our formulation
we must consider again  the spectral parameter independent case as in the
previous example and choose $\tilde Z=1$ and
 $Z_{12}\equiv ({A}_{12})^{-1}$.
This corresponds to   {\it case } 5 of sect. 5.1
and we  identify $B$ with $k^{\frac {1} {2}}$.
The relations (\ref{babcurrent1}) and  (\ref{babcurrent2}) are then recovered
directly from  (\ref{babcurrent10})
and the  braiding equations
(\ref{zlzl1+}--\ref{zlzl2+}).
For obtaining the monodromy relations we define $ Q_j\equiv  B^{-1}T^{[j,3]}$
 and   $ S\equiv  B^{-1}T$ and
repeating
the steps of the previous example
derive
 the   relations (\ref{babmono1}) and (\ref{babmono2}).
\subsection { Nonultralocal quantum mapping }

The quantum mapping
associated with the
lattice Gelfand-Dikii hierarchy is
given by the Lax operator $L_n$ in the form \cite{Nijhof}
$L_n=V_{2n}V_{2n-1}$, where
\[V_{n}=\Lambda_n \left(1+\sum^N_{i>j=1}v_{i,j}(n) e_{i,j}\right).\]
The commutation relation of the hermitean
operators $v_{i,j}$ is
\[ [v_{i,j}(n),v_{k,l}(m)]=
 h(\delta_{n,m+1}\delta_{k,j+1}\delta_{i,N}\delta_{l,1}-
 \delta_{m,n+1}\delta_{i,l+1}\delta_{k,N}\delta_{j,1})\]
 which aquires nontrivial values also at nearest-neighbour lattice points.
This clearly makes the corresponding Lax operator  nonultralocal in nature.
 The related $R$-matrix is given by the rational solution
 \be R_{12}(u_1-u_2)= {\bf 1}+ \frac { h P_{12}}{u_1-u_2}
.\ll{exR}\ee
It has been found in \cite{Nijhof} that the quantum
 Yang-Baxter equation for this
model may be given by
\begin{equation}
\tilde R_{12}(u_1-u_2)L_{1j}(u_1)L_{2j}(u_2)
= L_{2j}(u_2)L_{1j}(u_1){R}_{12}(u_1-u_2)
,\ll{nijhof1}\end{equation}
along with
nonultralocal conditions
\begin{equation}
 L_{2 j+1}(u_2)S_{21}(u_1)L_{1 j}(u_1)
=L_{1 j}(u_1)L_{2 j+1}(u_2)
\ll{nijhof3}\end{equation} and
\begin{equation}
 L_{2 k}(u_2)L_{1 j}(u_1)
=L_{1 j}(u_1) L_{2 k}(u_2), \ \
\ll{nijhof4}\end{equation}
 {for} $  k>j+1.$ In the above formulas

\be \tilde R_{12}(u)=
 (S_{12}(u_2))^{-1}  R_{12} (u_1-u_2) S_{21}(u_1)  \ \ \mbox {and }\ \
 S_{12}(u_2)
=
 {\bf 1} - \frac { h }{u_2}\sum_{\alpha}^{N-1}
 e_{N\alpha}\otimes e_{\alpha
 N}\ee

The equation for the monodromy matrix with the periodic
boundary condition is of  the form

\begin{equation}
\tilde R_{12}(u_1-u_2)T_{1}(u_1)S_{12}(u_2)T_{2}(u_2)
= T_{2}(u_2)S_{21}(u_1)T_{1}(u_1){R}_{12}(u_1-u_2)
.\ll{nijhof2}\end{equation}

This model actually goes beyond the scope of our formalism, because
the matrix  $ S_{12}$ depends on the spectral parameter
and besides that it does not satisfy the YBE (\ref{zzz}). Nevertheless,
if we put formally
 $\tilde Z=1$ and $Z_{12}\equiv S_{12}^{-1}(u_2)
, $ then  we can see that $\tilde R$
 is of the form (\ref{rrr1}), which was mentioned in  section 2 as
an alternative
 to (\ref{zzz}). Morerover this choice fits into our ansatz (\ref{case2})
 and
satisfies
the restriction (\ref{sym2}).
 Therefore the relevant
equations can be derived from  (\ref{bqybel2})- (\ref{bqybet2}).

\section { New nonultralocal quantum models through application
 of   braided
structures  }
\setcounter{equation}0
For showing the usefulness of
our scheme based on the braiding relation and the braided quantum
Yang--Baxter equation,
we present in this section
some new examples of  nonultralocal quantum models
constructed from them.

\subsection { Quantum mKdV model  }
mKdV equation
\begin{equation}
\pm v_t+v_{xxx}+12v^2v_x=0
\ll{mkdv}\end{equation}
is a well known classical integrable system  \cite{Zakharov} with wide range
of applications \cite {mkdv}. This model possesses an infinite set of
conserved quantities
 $c_n, \  n=1, 2, \ldots $ at the classical level and exhibits
noncanonical Poisson bracket structure
\begin{equation}
\{ v(x),v(y) \}=\mp 2\delta'(x-y)
.\ll{mkdvpb}\end{equation}
 (\ref{mkdv}) may be derived as a Hamilton equation from
 \[ H= c_3=\frac {1} {2} \int_{-\infty}^{\infty} dx \left( v_x^2+v^4 \right) \]
 by using    (\ref {mkdvpb}).
 However, as it is known,
due to its nonultralocal algebraic structure
 (\ref {mkdvpb}) the much awaited quantum generalisation
of this model  could not be achieved through usual proceedure \cite{Faddeev,
kul-skly}.

We  propose     a new quantum version of the mKdV model
and show that its exact integrability at the lattice level
can be  established through
 our scheme built  on  the braided quantum algebraic structures.
For this purpose we perform lattice regularisation by constructing the
quantum  Lax operator of the discrete  model as

\be
L_{k}(\zeta) = \left( \begin{array}{c}
e^{-\frac {i}{2} v^-_k}
\qquad  \ \
\frac {\Delta}{2} \zeta
e^{\frac {i}{2} v^+_k}
 \\
-\frac {\Delta}{2} \zeta
e^{-\frac {i}{2} v^+_k}
\qquad \
e^{\frac {i}{2} v^-_k}
         \end{array}   \right),
\ll{Lmkdv}\ee

with the commutation relations between the operators $ v^\pm_k$
as
\be
[v^\pm_k, v^\pm_l]=\mp i\hbar (\delta_{k-1,l}- \delta_{k,l-1})
,\ll{cr1}\ee
\be
[v^+_k, v^-_l]=i\hbar (\delta_{k-1,l}- 2\delta_{k,l}+\delta_{k,l-1})
.\ll{cr2}\ee

Note that at the continuum limit, i.e. at  $\Delta \rightarrow 0,$
the discrete operators
$v^\pm_k$ go to the quantum field $\Delta v^\pm(x) $ and the relation
(\ref{cr1}) reduces to the nonultralocal commutation relation
\be
[v^\pm(x), v^\pm(y)]=\pm i\hbar (\delta_{x}(x-y)- \delta_{y}(x-y))
.\ll{cr3}\ee
 It may be puzzling at the first sight that the number of fields has
become  double. Therefore we should note that this pair of similar
 fields simply answer to the different possible signs of the mKdV eqn.
 (\ref {mkdv}) and the dependence of one of them drops out from the
 Lax operator (\ref {Lmkdv}) at the continuum limit:
 \[ L_{k}(\zeta) \rightarrow {\bf I} + \Delta {\cal L}(x,\zeta)+
 O ({\Delta}^2 ) \]
recovering the known Lax operator of the continuum model:
\be
{\cal L}(x,\zeta)= \frac {i}{2} (-v^-(x) \sigma^3+\zeta \sigma^2)
\ee
corresponding to (\ref {mkdv}) with $+$ sign. For getting the other sign
the role of $v^\pm_{k}$ should be interchanged in  (\ref {Lmkdv}).
A simple similarity transformation $ A^{-1} {\cal L}(x,\zeta) A=
{\cal L}_{mkdv}(x,\zeta)$
gives further the Lax operator of the mKdV model in the well known
AKNS form $
{\cal L}_{mkdv}(x,\zeta)= \frac {i}{2} (\zeta \sigma^3-v^-(x) \sigma^1).$

We show now that for finding the quantum $R$-matrix and the quantum
Yang--Baxter equation associated with the mKdV Lax operator
 (\ref {Lmkdv}) we may start with the braiding relation (\ref{zlzl1u}).
 Using (\ref{cr1}),(\ref{cr2}) and assuming $\tilde Z=1$
 and $\zeta=e^{i \lambda}, \eta=e^{i \mu} $ we find from

\begin{equation}
L_{2 j}(\lambda) L_{1 j+1}(\mu)
= L_{1 j+1}(\mu)Z_{12}^{-1}L_{2 j}(\lambda )
\ll{brmkdv}\end{equation}
the matrix
\be Z_{12}=  Z_{21}= q^{-\frac {1}{2} \sigma^3\otimes \sigma^3
},\ll{Zmkdv}\ee
 where $q=e^{\frac {i\hbar}{2}}.$
 Pluggging
 this $Z$-matrix in the braided QYBE (\ref{bqybel}) and using relations like
 \[ [e^{i\alpha v^{\pm}_j},  e^{i \beta  v^{\pm}_j}]=0, \quad
 e^{i \alpha v^+_j} \ e^{i \beta v^-_j}=e^{i 2 \alpha \beta \hbar }
e^{i \beta  v^-_j} e^{i\alpha v^+_j} \]

we derive after some algebra the quantum $R$-matrix of this model, which turns
out to be the standard trigonometric solution (\ref{R(u)}) for $m+n=2$:

 \be R(\lambda-\mu)=\sum_{\alpha,\beta=1}^2\left(\sin (\lambda-\mu)
e_{\alpha \alpha}\otimes e_{\beta \beta} +
\sin( (\lambda-\mu)+\frac {\hbar}{2})
e_{\alpha \alpha}\otimes e_{\alpha \alpha}
 + \sin \frac {\hbar}{2}
e_{\alpha \beta}\otimes e_{\beta \alpha}\right)\ll{Rmkdv}\ee
It is an easy check that (\ref {Rmkdv}) and (\ref{Zmkdv}) thus found
satisfy the prescribed consistency equations
({\ref{ybe}-\ref{zzru}) and (\ref{zzz}). Therefore, we may use
these $R,Z$-matrices to find the braided QYBE
for the monodromy matrix of the periodic quantum mKdV model
in the form (\ref{bqybet}). Moreover, due to the fact that the $R$-matrix
(\ref{Rmkdv}) commutes with  the $Z$-matrix (\ref{Zmkdv})
the braided QYBE's in the present case
simplify further to
\begin{equation}
{R}_{12}(\lambda-\mu) L_{1j}(\lambda)L_{2j}(\mu)
=  L_{2j}(\mu) L_{1j}(\lambda){R}_{12}(\lambda-\mu)
\ll{ybelmkdv}\end{equation}

and
\begin{equation}
{R}_{12}(\lambda-\mu) T_{1}(\lambda)Z_{12}^{-1} T_{2}(\mu)
=  T_{2}(\mu) Z_{12}^{-1} T_{1}(\lambda){R}_{12}(\lambda-\mu)
,\ll{ybetmkdv}\end{equation}
respectively. For identifying the commuting set of conserved quantities
from the trace identity,
 one may define     a operator valued matrix
 $ \hat l=e^{-\frac {i}{2} q_{N} \sigma^3}$, with $q_N$ canonical to
$v^\pm_N: [q_N,v^\pm_N]=  \pm i \hbar,$ so that
 $\hat T(\lambda)=\hat l^{-1} T(\lambda) \hat l. $
 Following the idea of sect. 5.2   one shows that $[tr \hat T(\lambda),
 tr \hat T(\mu)]=0$  and thus establishes the integrability relations
 of this
nonultralocal
 model at the quantum level.
 A more detailed account of this model including the Bethe ansatz
 solution will be given elsewhere.

\subsection {Quantum integrable anyonic supersymmetric model}
We have seen in sect. 6.2 that the SUSY models are covered by the
nonultralocal integrability scheme developed here. In this subsection we
propose background for
a new quantum integrable anyonic generalisation of the SUSY
models and show that it can be constructed in a systematic way
following the same formalism.

In analogy with the standard SUSY model we take $Z=\tilde Z$ but
choose its explicit form in a more general way:
  \be Z_{12}=\sum_{\alpha,\beta}
e^{i\theta \ \hat \alpha \hat \beta}
e_{\alpha \alpha}\otimes e_{\beta \beta}
\ll{Zany}\ee
where $\theta$ is the arbitrary anyonic phase.
For such a $Z$-matrix the braiding
relation describing the commuting relations between elements of the
Lax operator at different
lattice points $k>j$ take the form
\begin{equation}
L_{b_2(k)}^{a_2}(\mu) L_{b_1( j)}^{a_1}(\lambda)
= e^{-i \theta (\hat a_1- \hat b_1)(\hat a_2- \hat b_2)}
  L_{b_1( j)}^{a_1}(\lambda) L_{b_2(k)}^{a_2}(\mu)
,\ll{brany}\end{equation}
with all matrix indices running from $1$ to $N=m+n$.
The algebraic relations at the same point $l$ on the other hand
are given by the
corresponding braided QYBE
\begin{equation}
\sum_{\{k\}}R^{k_1k_2}_{a_1a_2}(\lambda-\mu)
L_{k_1(l)}^{b_1}(\lambda) L_{k_2(l)}^{b_2}(\mu)
e^{i \theta \hat k_2(\hat b_1- \hat k_1)}
=\sum_{\{k'\}}
e^{i \theta \hat a_1(\hat k'_2- \hat a_2)}
  L_{a_2(l)}^{k'_2}(\mu)L_{a_1(l)}^{k'_1}(\lambda)
R^{b_1b_2}_{k'_1k'_2}(\lambda-\mu)
,\ll{bqybelany}\end{equation}

  The quantum $R$-matrix may be chosen in the trigonometric form (\ref{R(u)})
  or as its rational limit
  \be
  R_{12}(\lambda)=E_{12}\lambda+\frac {\hbar}{2}P_{12}
  ,\ll{Rany}\ee

where $P$ is the permutation matrix and $E={\bf I}$ for the 'standard'
and $E_{12}= \eta_{12}$ for the 'nonstandard' solutions.
The consistency equations for $Z$ and $R$ matrices obviously hold for
all these solutions.
Taking the Lax operator of the associated
models related to the rational case as
\be
L_{a(l)}^{b}(\lambda)
=\lambda \delta_{ab} \ p^{0(l)}_{b}+ \frac {\hbar}{2}
e^{i\theta (\hat a \hat b)} p^{(l)}_{ba}
\ll{Lany}\ee
and using the  rational $R$-matrix  solution
(\ref{Rany}) one may derive from (\ref{bqybelany} )
the anyonic  super algebra (ASA) at the same lattice point by
matching the coefficients of different powers of the spectral
parameter. This yields the following set of algebraic relations
between the generators of this  graded algebra.
\be
 p^{(l)}_{b_1a_1}    p^{(l)}_{b_2a_2}-
 e^{i\theta (\hat a_1 \hat b_2-
\hat b_1 \hat a_2) }
 p^{(l)}_{b_2a_2}    p^{(l)}_{b_1a_1}=
  e^{-i\theta (\hat a_1 \hat a_2
)}\delta_{a_1b_2}  p^{(l)}_{b_1a_2} p^{0(l)}_{a_1}
- e^{i\theta (\hat a_1 \hat b_2-
 \hat b_1 (\hat a_2+\hat b_2)) }
\delta_{b_1a_2}  p^{0(l)}_{b_1} p^{(l)}_{b_2a_1}
\ll{alg1any}\ee
and
\be
  p^{0(l)}_{a_1}   p^{(l)}_{b_2a_2}=
 e^{i\theta (\hat a_1( \hat b_2-
 \hat a_2)) }
p^{(l)}_{b_2a_2}    p^{0(l)}_{a_1}
\ll{alg2any}\ee

for the standard $R$-matrix. In the nonstandard case
one obtains another type of ASA, where additional
multiplicative factors  $
e^{i\pi  (\hat a_1 \hat a_2)},\quad$ and $\ \
e^{i\pi  (\hat b_1 \hat b_2)},\quad$
appear  before the first and the second term of algebra (\ref{alg1any}),
respectively, while the relation  (\ref{alg2any}) turns into
\be
  p^{0(l)}_{a_1}   p^{(l)}_{b_2a_2}=
 e^{i(\theta+\pi) (\hat a_1( \hat b_2-
 \hat a_2)) }
p^{(l)}_{b_2a_2}    p^{0(l)}_{a_1}
\ll{algns2any}\ee
In the nonstandard case, as can be easily checked,  we obtain also an extra
relation
\[(p^{(l)}_{ab})^2=0, \quad \mbox {for} \quad \hat a-\hat b=1,\]
reflecting the fermionic type character.
We use  braiding
relation  (\ref {brany}),
for obtaining the algebra  at different lattice points:

\be
 p^{(k)}_{a_2b_2}    p^{(j)}_{a_1b_1}=
 e^{-i\theta (\hat a_1- \hat b_1)
(\hat a_2- \hat a_2) }
 p^{(j)}_{a_1b_1}    p^{(k)}_{a_2b_2}
\ll{alg3any}\ee
for $k>j$ with all other commutators being trivial. This relation remains
same in both standard as well as nonstandard cases, since the baiding
relation does not involve $R$-matrix solutions.

 Note that this algebra is a
 nontrivial generalisation of the  well known graded super algebra \cite{SUSY},
where together with an arbitrary anyonic parameter $\theta$ a set of
$N$ additional operators $  p^{0(l)}_{a}    $  appear. For recovering
the known super algebra we have to start with our relations corresponding
to the nonstandard $R$-matrix and choose  $ \theta=\pi$. It is easily seen
from (\ref {algns2any}) that for such a choise  $  p^{0(l)}_{a}    $
commute with all other operators and hence may be chosen as unity.

For constructing quantum integrable models involving $m$ number of
bosons $b^{(k)}_a$
and $n$ number of
anyons $f^{(k)}_\alpha $,
as a direct generalisation of the integrable SUSY models,
we may consider a realisation of the generators as
 \be
p^{(k)}_{ab} = b^{\dagger(k)}_a \ b^{(k)}_b, \quad
p^{(k)}_{\alpha a} = f^{+(k)}_\alpha \ b^{(k)}_a,\\  \quad
p^{(k)}_{a \alpha } = b^{\dagger(k)}_a \ f^{(k)}_\alpha,\quad
p^{(k)}_{\alpha \beta} = f^{+(k)}_\alpha \ f^{(k)}_\beta,
.\ll{realany}\ee
The bosons satisfy the standard  commutation rules  and they
commute with all anyons, while  the anyonic operators have
the following commutation relations involving the phase factor $\theta$.
\be
 f^{(k)}_\alpha \ f^{+(j)}_\beta = e^{i\theta} \
 f^{+(j)}_\beta \ f^{(k)}_\alpha ,\ll{any1}\ee
 \be
 f^{(k)}_\alpha \ f^{(j)}_\beta = e^{-i\theta} \
 f^{(j)}_\beta \ f^{(k)}_\alpha ,\ll{any2}\ee
for $k>j$ and \be              [ p^{0(k)}_{\alpha },  \
 f^{(j)}_\beta]=0
  ,\ll{any2a}\ee
for $k\neq j$, while
\be
 f^{(l)}_\alpha \ f^{+(l)}_\beta - e^{i\theta} \
 f^{+(l)}_\beta \ f^{(l)}_\alpha = \delta_{\alpha \beta}
 p^{0(l)}_{\alpha }  \ll{any3}\ee
and
\be
p^{0(l)}_{\alpha }   f^{(l)}_\beta =  e^{-i\theta} \
 f^{(l)}_\beta p^{0(l)}_{\alpha }
, \quad [ f^{(l)}_\alpha ,\ f^{(l)}_\beta]=0,
  \ll{any4}\ee

at the same lattice points. Note that these relations are consistent
with the anyonic algebra well known in the literature \cite{anyon0,anyon1}.
The existence of such anyonic
operators in one dimension is also  in accordance with
 \cite{anydim}, where the existence of
 anyons has been established in any dimensions.
 It is important to note  that though  one can
take $ f^{+(k)}_\alpha = ( f^{(k)}_\alpha)^\dagger $
to derive other commutation
 relations for $k \neq j$, it seems  not to hold
 at the same lattice ponts with nontrivial $\theta$.
This   kind  of problem arising at the coinciding points also  known to exist
 in the  standard anyonic theory
 \cite{anyon2,anyon1}. The associated anyonic SUSY integrable models
 should be represented
by the Lax operator (\ref{Lany}) along with  realisations
like  (\ref{realany}).
An explicit construction of such a  model
 and its exact solution will be presented
in a future  publication \cite{future}.

\subsection {$q$-deformed anyoinic super algebra}

We may generate a novel $q$-deformation of the above  ASA (\ref {alg1any})-
-(\ref {alg3any}) by simply taking the same $Z$-matrix (\ref {Zany})
and choosing the trigonometric $R$-matrix solution (\ref {R(u)}). Representing
the Lax operator through the corresponding generators $
\tau^{\pm(l)}_a, \tau^{(l)}_{ab}$ of the algebra
as
\be
L_{a(l)}^{b}(\lambda )
= \delta_{ab} \ (e^{i\lambda}  \tau^{+(l)}_{b}+e^{-i \lambda}
\tau^{-(l)}_{b})
+
e^{i\theta \ \hat a \hat b}\tau^{(l)}_{ba}
\ll{Lqany}\ee
 and comparing the coefficients of different spectral parameters
 from (\ref{bqybelany} ) and (\ref{brany} )
 we derive the set of relations at coinciding
 as well as different points forming  the $q$-ASA. For example, a set
 of such relations in the general form  for arbitrary $m,n$ grading
 looks like
\bea
& & e^{-i\theta (\hat a_1 \hat c_2)}\left ( (q^{f(a_1)}-1)
 \delta_{a_1a_2} + 1 \right)\tau^{+(l)}_{c_1a_1} \tau^{+(l)}_{c_2a_2}
+(q-q^{-1})
e^{-i\theta (\hat c_2  \hat a_2)}
\tau^{+(l)}_{c_1a_2} \tau^{+(l)}_{c_2a_1}\ (a_1<a_2) \nonumber \\
&=&
e^{-i\theta (\hat a_2 \hat c_1)}\left ( (q^{f(c_1)}-1)
 \delta_{c_1c_2} + 1 \right)\tau^{+(l)}_{c_2a_2} \tau^{+(l)}_{c_1a_1}
+(q-q^{-1})
e^{-i\theta (\hat c_2  \hat a_2)}
\tau^{+(l)}_{c_1a_2} \tau^{+(l)}_{c_2a_1}\ (c_2<c_1)
\ll{qasa1G}\eea
and
\be
\tau^{+(k)}_{c_1a_1} \tau^{+(j)}_{c_2a_2}
=e^{-i\theta (\hat c_1- \hat a_1)(\hat c_2- \hat a_2)}
         \tau^{+(j)}_{c_2a_2} \tau^{+(k)}_{c_1a_1}
,\ee
where we have denoted $ \tau^{+(j)}_{aa}=\tau^{+(j)}_{a}, \quad
\tau^{+(j)}_{ca}=\tau^{(j)}_{ca},$ for $\ c>a \ $ and $f(a)=1$ for  the
standard $R$-matrix, while
$f(a)=1+\hat a(\frac
{\pi}{\eta}-2)$ for the nonstandard solution.

For simplicity we present here the $m=1, n=1$ case corresponding to the
standard trigonometric $R$-matrix solution in the
following explicit form.
\bea
\tau^{(l)}_{21} \tau^{(l)}_{12} &-&
e^{-i\theta }
\tau^{(l)}_{12} \tau^{(l)}_{21}= (q-q^{-1})
   \left(
   \tau^{+(l)}_{1} \tau^{-(l)}_{2}-  \tau^{-(l)}_{1} \tau^{+(l)}_{2}
\right)\nonumber \\
 \tau^{\pm(l)}_{a} \tau^{(l)}_{ab}&=& q^{\pm} e^{i\theta \hat a}
 \tau^{(l)}_{ab}  \tau^{\pm(l)}_{a}, \quad
 \tau^{\pm(l)}_{a} \tau^{(l)}_{ba}= q^{\mp 1} e^{-i\theta \hat a}
 \tau^{(l)}_{ba}  \tau^{\pm(l)}_{a}
,\ll{qasa1}\eea
at the same lattice point $l,$ with all $ \tau^{\pm(l)}_{a}$  mutually
commuting and $a,b =1,2.$
The nontrivial  commutation relations  at different points are given by
\be
 \tau^{(k)}_{12}   \tau^{(j)}_{21} =
 e^{i\theta }  \tau^{(j)}_{21}   \tau^{(k)}_{12}, \quad
 \tau^{(k)}_{ab}   \tau^{(j)}_{ab} =
 e^{-i\theta }  \tau^{(j)}_{ab}   \tau^{(k)}_{ab},
\ll{qasa2}\ee
for $k>j$. This super algebra
represents  a  generalisation of the extended trigonometric
Sklyanin algebra \cite {kundumallick} to include an anyonic parameter
$\theta$ and
nonultralocal algebraic property  (\ref{qasa2}).
This algebra would naturally lead to a novel anyonic quantum group
with  'long range' nonultralocality.
One can recover the super $sl_q(m.n)$ algebra introduced in
\cite{SUSYq}
by choosing $\theta=\pi$ and considering the nonstandard $R$-matrix.
Realisation of this algebra through anyonic quantum group  as well as
through a new anyonic $q$-oscillator is possible. Detailed discussion
of this problem along with concrete
model construction will be given elsewhere
\cite {future}.

\section{Conclusion}
 \setcounter{equation}0
We have presented a scheme suitable for describing quantum nonultralocal
models including supersymmetric models. The scheme is based on the concept
of quantised braided groups. We have noticed that while the braiding
of supersymmetric models is 'uniform', i.e. the commutation relations of spatially
separated objects remains the same independent of their distance, the
commutation relations in nonultralocal models exhibit different properties
for near and
 distant neighbours.

 For that reason we generalise the quantised braided
 groups to include  different types of braiding to distinguish between
nearest and nonnearest neighbours, and express it in the
Faddeev--Reshetikhin--Takhtajan algebra form.
To frame the formalism for application to integrable models we introduce
spectral parameter through Baxterisation of the
Faddeev--Reshetikhin--Takhtajan relations. This
results in a braided quantum Yang--Baxter equation
 along with nonultralocal commutation
relations initiated by the braiding. Due to the nonultralocal
commutation relations,    the underlying coalgebra structure
remains intact and that in turn enables us to construct monodromy matrices
satisfying braided quantum Yang--Baxter equation
 both for periodic and finite lattices.

In contrast to ultralocal models, the problem of deriving a set of
commuting operators from quantum Yang--Baxter equation
 becomes nontrivial.
Nevertheless we are able to solve it for several classes of braiding.

We show the scope of our formalism by describing the supersymmetric
 as well as
nonultralocal quantum models proposed in the context of  integrable or
conformal field theory,
 as different examples of the theory presented here. The last
 known example we consider
does not fit completely in our scheme and suggests some possibilities
of its further generalisation to spectral parameter
dependent braidings.

As further  applications of the proposed scheme
we are able to
find new examples of nonultralocal models as quantum integrable systems.
Following our formalism we have solved the  important
problem of describing the well known
classically integrable nonultralocal  mKdV model
at the quantum level.
We have  shown  the possibility of constructing a new class
of SUSY models involving bosonic  as well as anyonic
fields. Finally we find a novel quantum deformation of the anyonic super
algebra with arbitrary grading.

It would be interesting also to
 explore
various  other possibilities
by taking the  braiding
as suggested in sect. 6.1,
apart from  the long-standing problems of
 quantising   nonultralocal models like
nonlinear $\sigma$-model, derivative nonlinear Schr\"odinger equation,
 sine-Gordon
equation in light--cone coordinates etc.
For inclusion of wider class of nonultralocal models, e.g. KdV model
in its second Hamiltonian formulation related to the Virasoro algebra,
it seems that one would need  braided structures containing more than
two types of braiding. This offers another direction for extension
of the present formalism.
 \\ \\
 {\it Acknowledgement}\\
One of the authors (AK) acknowledges the support of Alexander von Humboldt
Foundation research fellowship grant and expresses his thanks to Prof.
Vladimir Rittenberg, Fabian Essler and other members of the Theory Division
of the Physikalisches Institut, Bonn for stimulating discussions.

L. Hlavaty would like to express his gratitude to Physikalisches
Institut, Bonn for hospitability during the preparation of the final
version of the paper and acknowledges the support of the grant
No. 202/93/1314 of the Czech Republic.

\end{document}